# SAMPA Based Streaming Readout Data Acquisition Prototype

E. Jastrzembski
Physics Division
Thomas Jefferson National Acelerator Facility
Newport News, Virginia, USA
jastrzembski@jlab.org

D. Abbott
Physics Division
Thomas Jefferson National Accelerator Facility
Newort News, Virginia, USA
abbottd@jlab.org

J. Gu
Physics Division
Thomas Jefferson National Accelerator Facility
Newport News, Virginia, USA
jgu@jlab.org

V. Gyurjyan
Scientific Computing Division
Thomas Jefferson National Accelerator Facility
Newport News, Virginia, USA
gurjyan@jlab.org

G. Heyes
Scientific Computing Division
Thomas Jefferson National Accelerator Facility
Newport News, Virginia, USA
heyes@jlab.org

B. Moffit
Physics Division
Thomas Jefferson National Accelerator Facility
Newport News, Virginia, USA
moffit@jlab.org

E. Pooser
Advanced Concepts Labotatory
Georgia Tech Research Institute
Atlanta, Georgia, USA
Eric.Pooser@gtri.gatech.edu

C. Timmer
Scientific Computing Division
Thomas Jefferson National Accelerator Facility
Newport News, Virginia, USA
timmer@jlab.org

A. Hellman
Physics Department
University of Virginia,
Charlottesville, Virginia, USA
afh5rb@virginia.edu

*Abstract*—We have assembled a small-scale streaming data acquisition system based on the SAMPA front-end ASIC. We report on measurements performed on the SAMPA chip and preliminary cosmic ray data acquired from a Gas Electron Multiplier (GEM) detector read out using the SAMPA.

*Keywords— SAMPA, GEM, streaming, data acquisition*

## I. Introduction

We have assembled a small-scale streaming data acquisition system based on the SAMPA front-end ASIC. The goals of the prototype system are to determine if the SAMPA chip is appropriate for use in detector systems at Jefferson Lab, and to gain experience with the hardware and software required to deploy streaming data acquisition systems in nuclear physics experiments.

## II. SAMPA ASIC

The 32-channel SAMPA ASIC is a mixed signal IC designed for the high luminosity upgrade of the ALICE experiment at the CERN LHC (Fig. 1) [1] [2]. For each channel the analog front end of the chip is composed of a charge sensitive amplifier (CSA) followed by two shaping circuits that produce a $4^{th}$ order semi-Gaussian pulse. The amplifier gain can be programmed to 20 or 30 mV/fC. The shaped signal with a nominal peaking time of 160 ns is digitized by a dedicated 10-bit ADC operated at up to 20 MSPS (Fig. 2).

After leaving the ADC the digitized data can follow two alternative paths: be processed by a dedicated digital signal processor (DSP mode) or bypass the DSP and be sent out in direct ADC mode (DAS mode). The DSP can be utilized to perform baseline corrections and data compression (zero suppression, Huffman coding) before sending the processed data off chip in multiple serial data streams. In contrast, direct ADC mode sends out the unprocessed ADC data directly via ten serial data links. The design of the Front-End Card (FEC) that supports the SAMPA chips limits the sampling rate in DAS mode to 5 MSPS. The SAMPA chip can operate in both streaming and triggered modes.

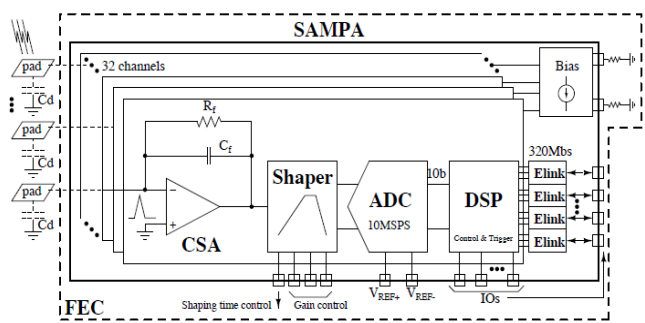

Figure 1. SAMPA integrated circuit

The following describes SAMPA operation in streaming DSP mode with zero suppression applied. A programmable processing window measured in number of ADC samples is defined (maximum 1024). In streaming mode when a processing window ends another immediately begins. If an input pulse that crosses the programmable threshold occurs in a processing window the SAMPA DSP unit constructs a *packet* of data consisting of the raw ADC samples of the shaped pulse and the time of the 1st sample. Up to 3 pre-samples (samples before pulse initial threshold crossing) and 7 post-samples (samples after pulse return below threshold) may be included. Similar data from additional pulses within the window are appended to the same packet. When the processing window ends the packet data is transported off chip on serial data lines (e-links). Data from the 32 input channels are shared on up to 11 serial e-links.

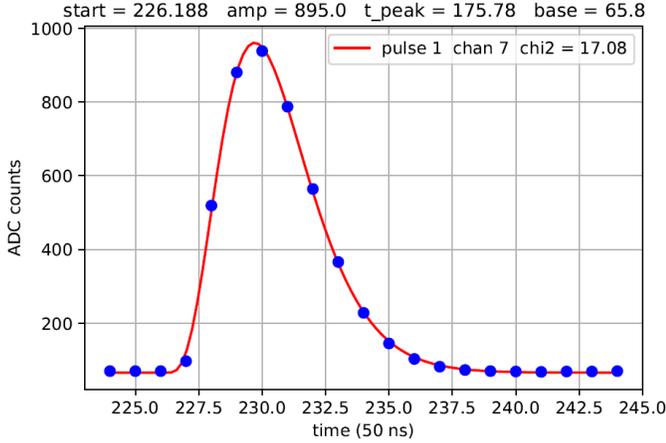

Figure 2. Fit to shaped pulse sampled at 20 MSPS

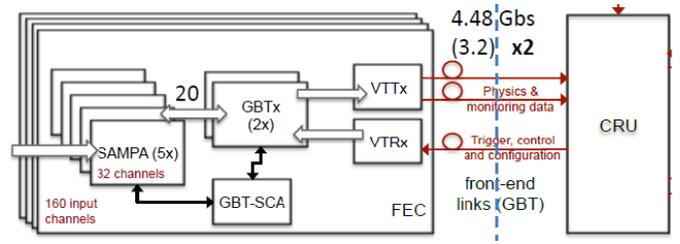

Figure 3. Readout system

### III. READOUT

The 800 channel system is composed of components used in the ALICE Time Projection Chamber (TPC) data acquisition upgrade (Fig. 3). Five front-end cards support five SAMPA chips each. SAMPA data streams on an FEC are concentrated into two high-speed (4.48 Gbs) serial data streams by a pair Gigabit Transceiver ASICs (GBTx) [3] [4]. These ten streams (44.8 Gb/s) are transmitted from the FECs over fibers to a PCIe based Common Readout Unit (CRU) [5]. The FPGA engine on the readout unit filters and compresses data that the host PC transmits to a server via 100 Gbs ethernet. In addition to the uplink paths used for data and monitoring, the GBTx provides a fixed latency downlink path for trigger, control, and configuration of FEC components. Components on the FECs are radiation tolerant. High data rates can be handled. The system is by design scalable.

### IV. MEASUREMENTS

We performed fundamental measurements on the SAMPA ASIC in streaming mode. These were done with controlled test pulses at the SAMPA inputs. A negative voltage step V is applied to a 1 pF capacitor (2% tolerance) that is connected to the SAMPA input. The leading edge is fast (~4 ns) so that we can observe the system's response to an impulse stimulus. The trailing edge is very slow (~25 us) so that the reverse current pulse is virtually undetectable in measured amplitude. The charge injected is Q = CV. In some studies we have purposely slowed the leading edge of the voltage step to simulate the injection of the given charge over a longer time period (as happens in real detectors). Pulses in our study were generated by a precision pulse generator (Tektronix AFG3252C) and then attenuated by a factor of 10.

When applying external pulses to a synchronous system there is concern that systematic biases could be introduced due to the time relationship of the pulses to the relatively coarse (50 ns) sampling clock. Generating a large number of pulses at random times assures that such systematic effects are properly averaged. However, we chose a different approach that allows us to study these systematic effects while averaging over them. The study also enables the possibly of correcting for the systematic effects on a hit by hit basis in the data stream.

We set a SAMPA processing window size of 1000 ADC samples (50.000 µs for 20 MSPS) and apply external pulses regularly with a period of 50.004 µs. This results in one hit per processing window, greatly simplifying data analysis. Once the initial phase relationship of pulser to sampling clock is set, consecutive pulses are each shifted 4 ns later in relation to the clock. In 25 pulses the original phase relationship of the pulser to sampling clock is recovered. We are effectively exploring systematic timing effects every 2 ns. As long as our data run has N*25 pulses (N = 1,2,3,..), any such systematics will be properly averaged. Typical runs we take are 200 or 400 pulses. We have studied how the pulse fit parameters of amplitude, peaking time, start time, and the computed integral depend on the pulse to sampling clock relationship. The periodicity of the effects is confirmed to be 25 pulses as expected. The duration of the run is short enough (<20 ms) that drift due to the use of independent time bases for the pulser and sampling clock is not an issue. The effects are seen to be nearly independent of input pulse amplitude. At 20 MSPS the effects are small (~0.5%) and can be neglected. At 10 MSPS the effects are significantly worse (~3%). We are working on an algorithm that will correct for these effects. The corrections are based on where along the pulse shape the samples are taken. Once implemented we can perform the corrections to the pulse parameters on a hit by hit basis in the data stream.

All parameters are allowed to float in our pulse fits. If we fix the peaking time to its nominal value (160 ns) the above systematic effects in the remaining parameters are reduced to about 1% at 10 MSPS. However, in actual detectors charge may arrive at the readout pads over extended time periods, thus broadening the pulse (see section C below). So we feel the best approach is to let all fit parameters float and make corrections.

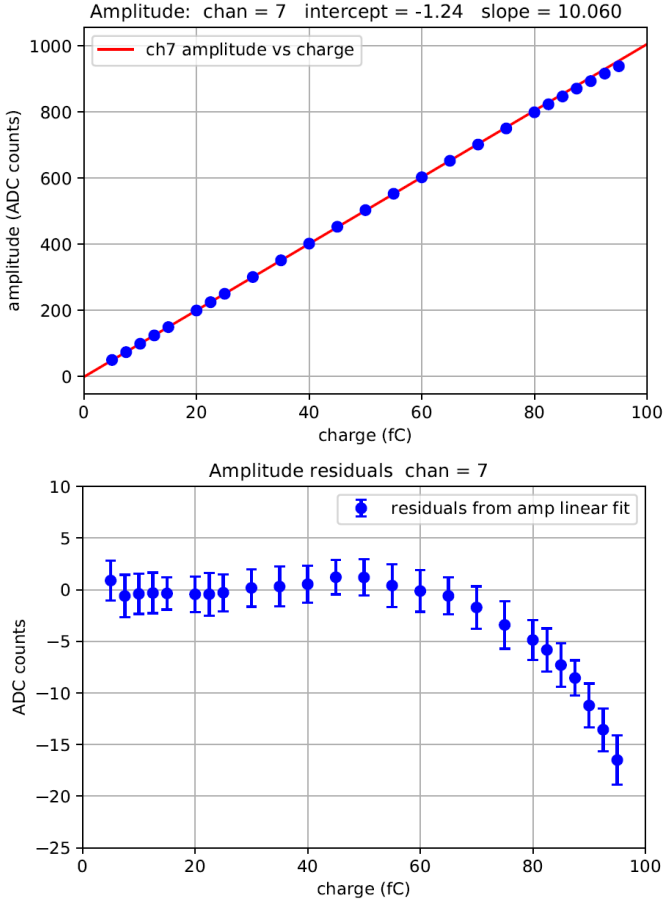

Figure 4. Pulse amplitude vs. Input Charge. Pulse amplitude is determined by fitting to the SAMPA impulse shape function. Linear fit is over 7 – 70 fC.

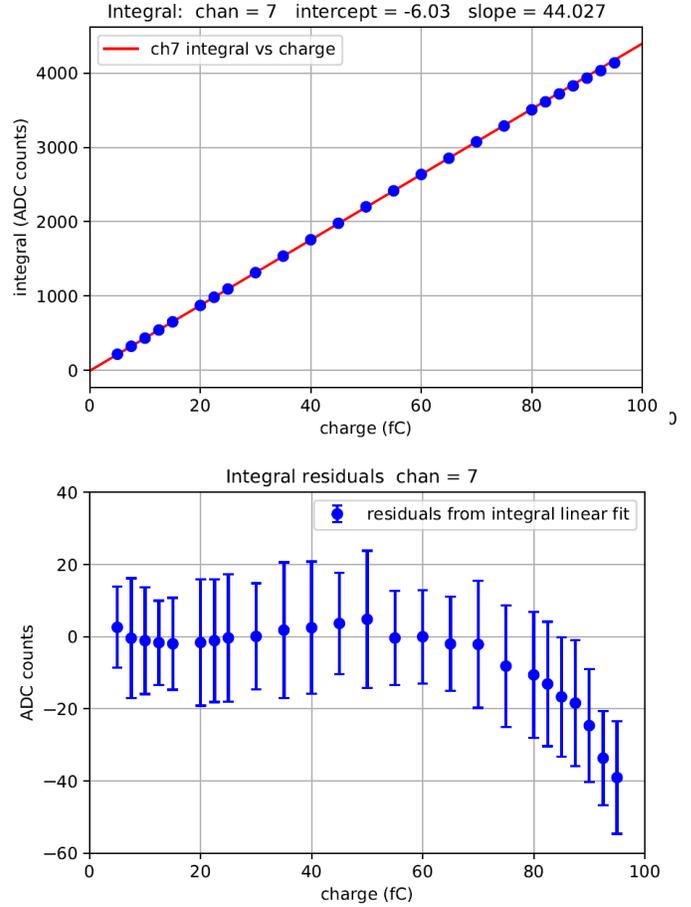

Figure 5. Pulse integral vs. Input Charge. Pulse integral is the sum of ADC samples of the pulse minus an estimate of the baseline sum beneath the pulse. The baseline sum estimate is derived from an average of samples before and after the pulse. Linear fit is over 7 – 70 fC.

*A. Linearity*

Using these controlled pulses we have performed linearity measurements at gains of 20 and 30 mV/pC, and with sampling rates of 10 and 20 MSPS. Measurements with a gain of 20mV/pC with 20 MSPS are shown in Fig. 4 and Fig. 5. The amplitude of the fitted 4th order semi Gaussian pulse is plotted on the vertical axis in Fig. 4. In the fits all parameters are allowed to float. Alternatively, a pulse integral is defined as the sum of ADC samples of the pulse minus an estimate of the baseline sum beneath the pulse. Pulse integral is plotted on the vertical axis of Fig. 5. In both linearity plots a saturation effect attributed to the shaping circuit is visible for charge greater than 75 pC. This can easily be corrected for in the data. We found that the saturation effect appears larger in 20 MSPS measurements than those at 10 MSPS. We are investigating this further.

*B. Time Resolution*

We have measured the time resolution of the SAMPA chip under ideal conditions using the following technique. Two pulses having a fixed time relationship are applied to separate inputs of a SAMPA chip. For each pulse the start time from the fit is taken as the pulse time. For each pulse pair the time difference $\Delta t$ is computed and time difference resolution $\sigma(\Delta t)$ is determined from the distribution of $\Delta t$. The time resolution $\sigma(t)$ of a hit is estimated as $\sigma(\Delta t) / 1.414$.

We expect the time resolution to depend on input pulse size, so data was collected for input pulse pairs with 5, 10, 20, 40, 60, 80 mV amplitudes (1 mV = 1 pC). The pulses of a pair are separated by > 600 ns to eliminate any influence due to crosstalk between the channels. The time resolution may also depend on the relationship of the pulses of the pair to the ADC sampling clock. To study this, data was collected for pulse delays of 600, 610, 620, 630, 640, 650 ns. This covers the entire ADC sampling interval of 50 ns. A total of 36 data runs were made for this study (6 amplitudes, 6 delays). Results are shown in Fig. 6.

Fig. 6a clearly shows that larger pulses have better time resolution, and that for these larger pulses the resolution does depend on the delay between pulses. Fig. 6b summarizes the results of 6a by taking the worst case resolution for each input pulse amplitude. It is at first surprising that with a sampling period of 50 ns, time difference resolutions of a few ns can be achieved. However, fitting the 14-20 samples of a pulse to the shape severely constrains the fitted start time parameter.

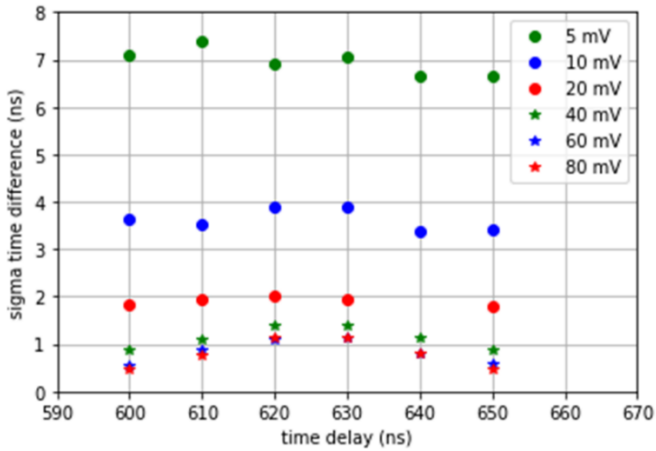

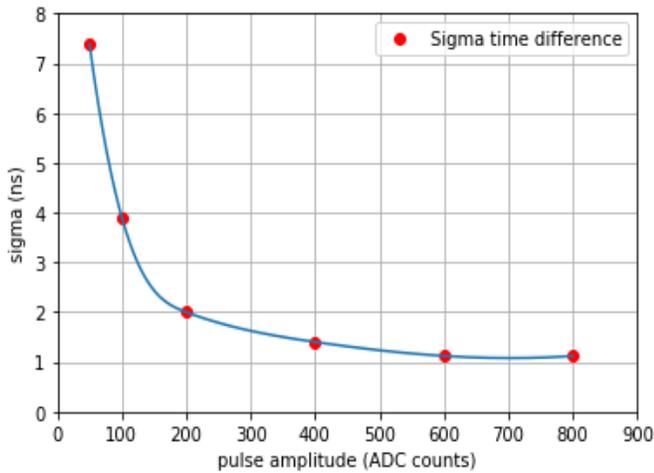

Figure 6. (a) Time difference resolution vs. time delay of pulse pair, for amplitudes across the entire dynamic range. Start time of the fitted pulse defines the pulse time. (b) Worst case time difference resolution vs. pulse amplitude (condensed from (a)).

We expect that for less than perfect pulses (i.e. from a real detector) the time resolution will degrade noticeably. We plan to study this with cosmic ray events in the GEM detector by comparing times on adjacent strips of a cluster as well as comparing times for the same hit on different GEM layers.

### C. Effect of Charge Injection Period

We have measured how the SAMPA pulse shape changes when charge arrives at its input over extended time periods. This is particularly relevant when reading out detectors like a TPC where the angle between the charged particle's momentum vector and the drift direction may be small. Fig. 7 shows a comparison of the impulse shape to the expected pulse shape when the same charge arrives uniformly over a period of 140 ns. In our test setup we can control the period over which the charge is injected by varying the transition time of the step pulse's leading edge. Because the voltage ramp is linear, the charge injection is uniform over time. We have measured the SAMPA response with pulse transition times of 4, 40, 80, 100, 120, and 140 ns. With V fixed at 80 mV, 80 fC of charge is injected in each case. Note that the times studied are less than the nominal pulse peaking time of 160 ns. In these cases a fit to the impulse shape with a larger shaping time yields a reasonable approximation to the measured pulse. Fig. 8 shows that the results agree well with a linear convolution model. Pulse broadening is accompanied by reduction of fit amplitude.

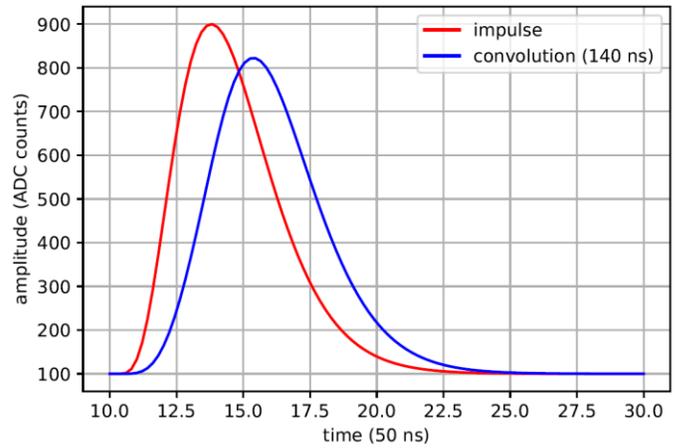

Figure 7. Predicted pulse shape (blue) for charge injected uniformly in time over a period of 140 ns from a convolution model. The red curve is the impulse response for comparison. The total charge (80 fC) and the start time is the same in both cases.

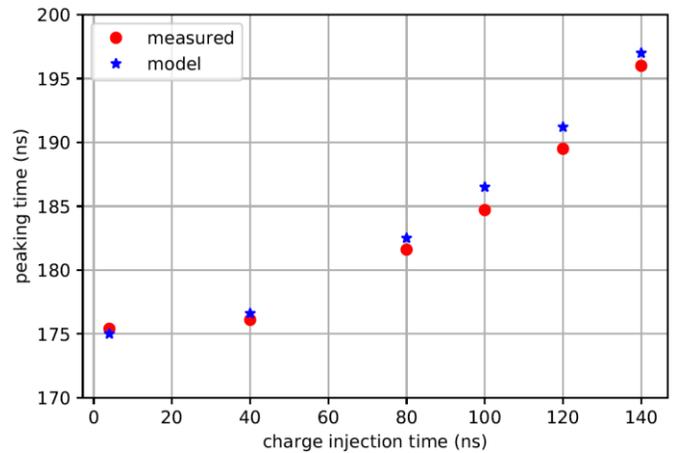

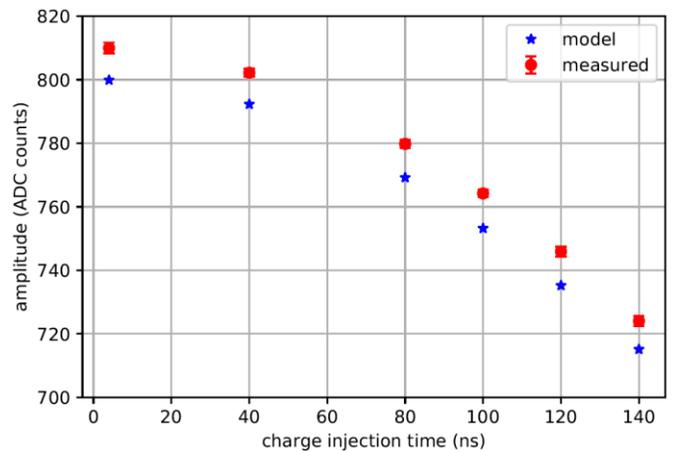

Figure 8. Comparison of peaking time and amplitude fits for convolution model and measured data at different charge injection periods. The fits are to the impulse shape function.

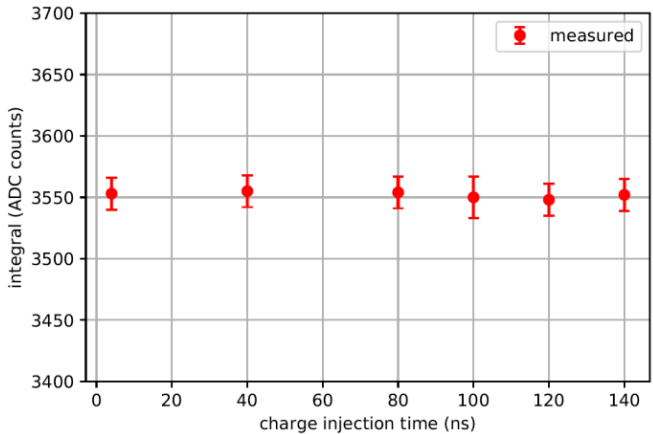

Figure 9. Computed pulse integral for measured data at different charge injection periods.

A correction to the pulse amplitude can be made based on the measured peaking time. Fig. 9 shows that the computed pulse integral is not sensitive to the charge injection time across the studied periods (4 – 140 ns).

An attempt was made to fit pulses with a convolution model shape function. A width parameter $W$ corresponding to the charge injection period was defined. The function was a summation of impulse functions each shifted in time by 1 ns extending across time $W$ and having equal weights $Q/W$. Although it could successfully fit some pulses it was sensitive to the parameter start values assigned in the fitting program. Appling the convolution fit to less than perfect pulses (i.e. from a detector) often resulted in a very poor and non-physical choice of parameters. So we feel that fits with the impulse shape function and a floating peaking time parameter to be more reliable.

Besides not being affected by the charge injection time the pulse integral has some additional advantages as a measure of input charge. The integral is less sensitive to the effects of abnormal pulse shapes that happen in real detectors than fitted pulse parameters. The simple algorithm for the pulse integral can easily be performed in the CRU's FPGA, so pulse feature extraction can be done in hardware to reduce data volume. Doing a nonlinear fit in an FPGA is much more difficult. If the integral is used instead of a fit, a pulse time must be estimated using samples on the rising edge of the pulse. Previous studies indicate that a resolution of about 1/3 of a sampling period (~16 ns here) can be achieved with this technique. This is worse than the time resolution obtained from a pulse fit.

*D. GEM Detector*

We have also coupled the readout system to a small Gas Electron Multiplier (GEM) detector and have begun to study the SAMPA's response to cosmic rays. The small prototype triple GEM has an active area of 153.6mm x 153.6mm with X and Y readout strips (400μm pitch, 80μm X strips, 340 μm Y strips). The 768 channels match well with our 800 channel SAMPA readout system. It was necessary to enclose the detector and cable assemblies within a Faraday cage to eliminate pickup of radiated EMI (Fig. 10). The pedestal noise of the system is consistent with the capacitance of the detector strips and cables that connect the detector with the front-end readout cards (Fig. 11). Fig. 12 shows amplitudes for a cluster of hits from a cosmic ray particle. Fig. 13 shows the cluster amplitude sum distribution in one GEM plane for a cosmic ray run. We are in the process of studying amplitude and timing correlations between the X and Y GEM readout planes.

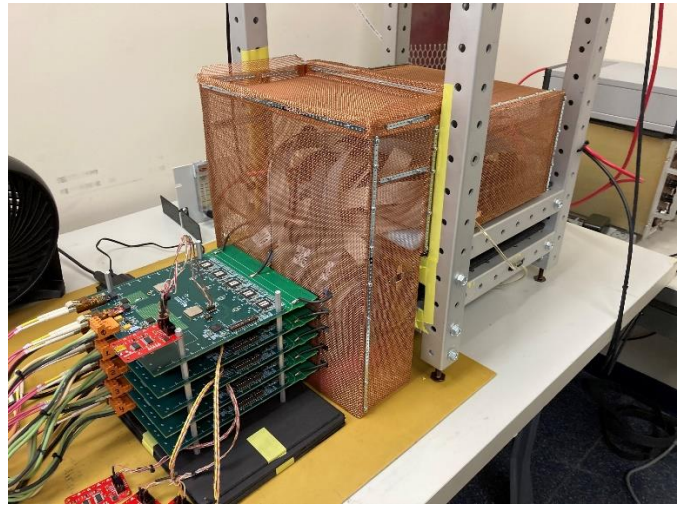

Figure 10. SAMPA front end cards connected to GEM detector. Detector and cables are enclosed in a Faraday cage.

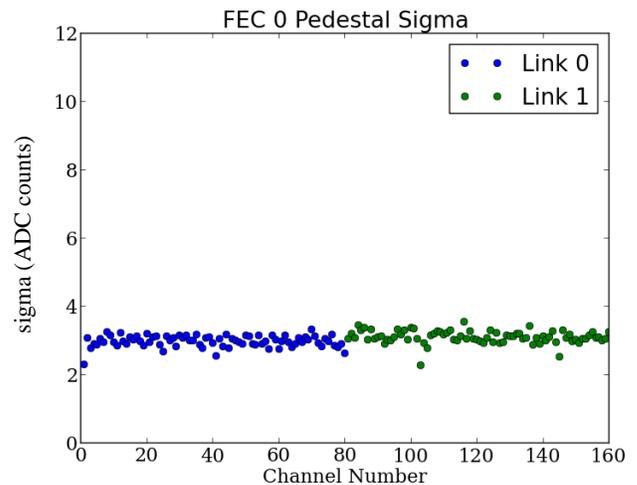

Figure 11. Pedestal noise (sigma) for detector system.

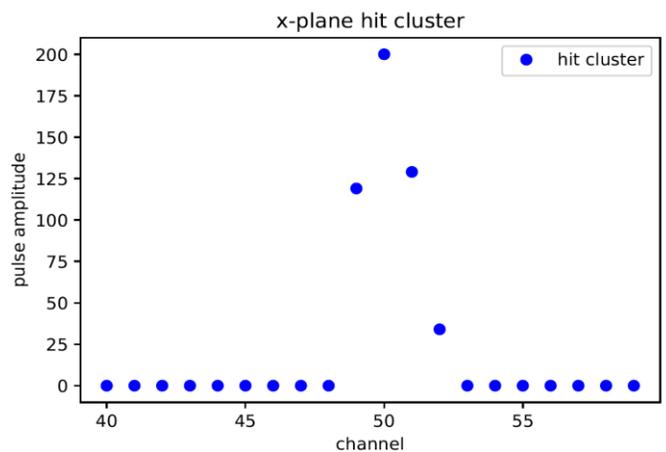

Figure 12. Cluster from a cosmic ray hit (sum = 482 (48 fC)).

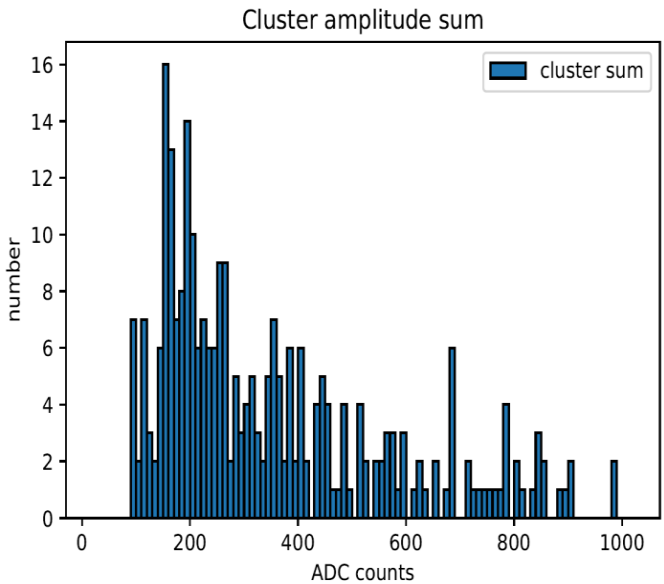

Figure 13. Cluster amplitude sum distribution in a GEM plane for cosmic ray events.

## V. CONCLUSION

We have successfully streamed and analyzed data from the SAMPA chip using both test pulse and GEM detector stimuli. We have made fundamental measurements on the SAMPA chip that complement those performed by the ALICE collaboration. We believe that the SAMPA chip and the elegant data transport mechanism employed in this system form an excellent basis for future streamed and triggered data acquisition systems at Jefferson Lab and the Electron-Ion collider.

## VI. FUTURE WORK

More extensive cosmic ray studies with the current readout system and GEM detector are planned. We will upgrade to a more powerful CRU [6] and will develop pulse feature extraction algorithms within the CRU's FPGA to reduce data volume. The upgraded system will serve as a test platform for streaming mode software development.

## ACKNOWLEDGMENT

We thank the following members of the ALICE collaboration for their help in getting our system up and running: Marco Bregant, Chuck Britton, Ken Read, Torsten Alt, Stefan Kirsch, and Anders Oskarsson. E.J. is especially grateful to Marco Bregant for generously sharing his expertise on the SAMPA chip with us.